\documentclass[journal,twoside,web]{ieeecolor}
\usepackage{lcsys}
\usepackage{cite}
\usepackage{amsmath,amssymb,amsfonts}
\usepackage{algorithmic}
\usepackage{graphicx}
\usepackage{textcomp}
\def\BibTeX{{\rm B\kern-.05em{\sc i\kern-.025em b}\kern-.08em
    T\kern-.1667em\lower.7ex\hbox{E}\kern-.125emX}}
\begin{document}
\title{Model Predictive Control with Preview: Recursive Feasibility and Stability}
\author{Xing Fang, \IEEEmembership{Member, IEEE}, Wen-Hua Chen, \IEEEmembership{Fellow, IEEE}
\thanks{This work was partially supported by the UK Engineering and Physical Sciences Research Council (EPSRC) Established Career Fellowship ``Goal-Oriented Control Systems: Disturbance, Uncertainty and Constraints'' under the grant number EP/T005734/1, the China Postdoctoral Science Foundation under Grants 2021M702505, and the 111 Project under Grants B12018. \emph{(Corresponding  author: Wen-Hua Chen)}}
\thanks{Xing Fang is with the Key Laboratory of Advanced Process Control for Light Industry of the Ministry
of Education, Jiangnan University, Wuxi, 214122, China (e-mail: xingfang@jiangnan.edu.cn) }
\thanks{Wen-Hua Chen is with Department of Aeronautical and Automotive Engineering, Loughborough University, Leicestershire, LE11 3TU, U.K. (e-mail: W.Chen@lboro.ac.uk)} }
\maketitle

\begin{abstract}
This paper proposes a stabilising model predictive control (MPC) scheme with preview information of disturbance for nonlinear systems. The proposed MPC algorithm is able to not only reject disturbance by making use of disturbance preview information as necessary, but also take advantage of the disturbance if it is good for a control task. This is realised by taking into account both the task (e.g. reference trajectory) and disturbance preview in the prediction horizon when performing online optimisation. Conditions are established to ensure recursive feasibility and stability under the disturbance. First the disturbance within the horizon is augmented with the state to form a new composite system and then the stage cost function is modified accordingly. With the help of input-to-state stability theory, a terminal cost and a terminal constraint are constructed and added to the MPC algorithm with preview to guarantee its recursive feasibility and stability under a pre-bounded disturbance. Numerical simulation results demonstrate the effectiveness of the proposed MPC algorithm.
\end{abstract}


\section{Introduction}
\label{sec:introduction}
\IEEEPARstart{I}{t} is well known that model predictive control (MPC), which is also known as receding horizon control (RHC), is a promising control technique based on on-line optimisation \cite{1,2,3}. However, the  presence of disturbance not only affects its performance but also may destroy the recursive feasibility and stability of an MPC algorithm. Depending on whether the information of disturbance is available for controller design or not, the techniques dealing with disturbance within the MPC framework broadly fall into two categories \cite{4}. 

The first category of MPC methods copes with disturbance under the assumption that the information of disturbance is not directly available. The nominal MPC algorithm utilizes its inherent robustness to attenuate disturbance, which ignores the disturbance in the on-line optimisation problem and MPC design \cite{5}. The nominal MPC in general is only able to cope with sufficiently small disturbance. A slightly large disturbance may degrade the control performance significantly, even lead to infeasibility and instability of the closed-loop system. Another approach is the so-called robust MPC method, which is developed based on the worst case of disturbance. In this type of MPC strategy, the tube-MPC attracts the most attention \cite{6,7,8}. The tube-MPC method consists of two parts: a nominal MPC controller to guarantee the desired performance for the undisturbed system, and a feedback controller to ensure the actual state to remain in the desired tube. Another impressive method is min-max MPC, which includes two optimisation problems \cite{9,10}. The dual stage cost in the min-max MPC method leads to heavy computational burden, which poses an obstacle for its engineering implementation. Furthermore, the stochastic MPC provides an alternative way to deal with stochastic disturbance \cite{11,12}. The use of statistic properties of disturbance makes it possible to achieve better control performance than the worst case-based MPC algorithms.

The other category of MPC methods attenuates the influence of disturbance by making use of the specific information of disturbance. This is driven by the advances in sensor technology and the development of the disturbance estimation techniques. For example,  LIDAR and cameras on intelligent vehicles can look ahead for certain distance, providing terrain and other traffic information \cite{13}. When the disturbance is not directly measurable or too expensive to measure, there has been significant progress in developing disturbance and uncertainty estimation techniques in the last two decades, please refer to \cite{14} for detail.  Within the MPC framework, a disturbance observer has been integrated to form a composite control scheme, known as disturbance observer-based model predictive control (DOB-MPC). The DOB-MPC method consists of two components: a disturbance compensation input based on the estimated disturbance by a disturbance observer to reject the disturbance, and an optimal control input to achieve the desired regulation or tracking performance. In \cite{15}, a DOB-MPC method is developed for a small-scale unmanned helicopter in the presence of wind gusts. A DOB-MPC method is proposed for a class of linear systems in the presence of disturbance and applied to a DC-DC buck converter system \cite{16}. More recently, instead of treating disturbance rejection and control completely separately, a disturbance rejection model predictive control (DRMPC) method is designed for a broad class of input-affine nonlinear systems with disturbance \cite{17}. In this approach,  the feedforward control component is designed first based on the disturbance estimation to compensate the matched disturbance, and the optimisation problem is then formulated to attenuate the residual disturbance, as well as to meet the required performance specifications. All these works aim to reject the influence of disturbance, rather than to make an attempt to take the opportunity that may arise due to disturbance.

Motivated by the observation that sometime disturbance may do a favour for realising a control task or goal,  a new type of MPC algorithms looks to exploit the information of disturbance further.  The preview information of incoming ocean waves is considered in optimal control for wave energy converters (WECs) to maximize energy output \cite{18}. A preview steering control algorithm with the future road curvature is developed for autonomous vehicles, which improves the tracking accuracy and steering smoothness \cite{19}. Recently, the preview information of disturbance has been used within a nominal MPC framework for a class of linear systems, which provides an alternative approach to exploit the preview information \cite{20}. Stability and recursive feasibility is established by the virtue of inherent robustness of the nominal MPC. Despite all the existing works, it still lacks of systematic and practical design and analysis tools to fully take the preview information of disturbance into account in the design of MPC controller, as well as to guarantee the recursive feasibility and stability of the closed-loop system under disturbance.

This paper presents a general MPC framework with disturbance preview for a class of nonlinear systems with stability guarantee. The proposed MPC automatically takes the opportunity of the disturbance preview information, or rejects disturbance, depending on its influence on the control task. This is achieved by integrating disturbance information and the reference information together, and performing a single online optimisation, rather than treating them separately (in parallel), e.g. \cite{15}  or sequentially (in series), e.g. \cite{17}. This makes it possible to further improve performance for the control system. First of all, an augmented system is obtained by combining the original nonlinear system with disturbance preview information. An MPC framework is then proposed for the augmented system, where the optimisation problem  is formulated with the preview information of disturbance naturally. To guarantee the recursive feasibility and stability of the closed-loop system under bounded disturbance, inspired by both the input-to-state stability (ISS) theory and the current MPC stability theory,  a terminal cost and terminal constraint are carefully constructed offline and added into the MPC algorithm with preview. 

The main contributions of this paper are concluded as below.

(1) The proposed MPC framework formulates an augmented system that contains the dynamics of the disturbance preview information. The preview information of disturbance is included in the optimisation problem, which is able to make full use of the disturbance to improve control  performance, including taking the predictable disturbance as the opportunity of the control system.

(2) The recursive feasibility and input-to-state stability is established for the closed-loop system in the presence of disturbance. An extra $\mathcal{K}_{\infty}$-function is adopted in deriving terminal elements, which increases the stability margin and strengthens the disturbance-rejection capability.

(3) The implementation issues for the proposed MPC framework are also discussed in this paper. The calculations of the terminal conditions are briefly discussed, which makes the proposed MPC method applicable.

The remainder of the paper is organized as follows. The problem formulation is given in Section 2. The MPC framework with the preview information of disturbance is proposed in Section 3. Section 4 discusses implementation issues about the proposed MPC. Section 5 presents  simulation results of the proposed MPC scheme with comparison of the existing MPC algorithms. Finally, conclusions are drawn in Section 6.

\emph{Notations:}  $\mathbb{R}$ and $\mathbb{I}$ denote the sets of reals and nonnegative integers. $\mathbb{R}^n$ denotes the $n$-dimensional Euclidean space. $\mathbb{I}_{[a,b]}$ denotes the set of integers from $a$ to $b$. A PC-set is a closed and bounded convex set containing the origin in the interior. For any $x \in \mathbb{R}^n$, ${\left\| x \right\|_P} = \sqrt {{x^T}Px} $ with $P \ge 0$. $\lambda_{\rm{max}} (P)$ and $\lambda_{\rm{min}} (P)$ denote the maximum and minimum eigenvalues of matrix $P$, respectively. A function $\alpha(\cdot)$: $\mathbb{R}_{\ge 0} \to \mathbb{R}_{\ge 0}$ is a $\mathcal{K}$-function, if it is continuous, strictly increasing and $\alpha (0) = 0$. A function $\beta(\cdot)$: $\mathbb{R}_{\ge 0} \to \mathbb{R}_{\ge 0}$ is a $\mathcal{K}_{\infty}$-function, if it is a $\mathcal{K}$-function and unbounded. $\alpha \circ \beta (\cdot) = \alpha(\beta (\cdot)) $ denotes the composition of two functions $\alpha$ and $\beta$.

\section{Problem Formulation}
\subsection{Controlled plant}

Consider the following discrete-time nonlinear system subject to persistent additive bounded disturbance
\begin{eqnarray}
x(k+1) =  f(x(k),u(k),w(k)),
\end{eqnarray}
where $x(k) \in \mathbb{R}^n$, $u(k) \in  \mathbb{R}^m$,  and $w(k) \in  \mathbb{R}^q$ denote the state, input and disturbance at the current time, respectively. $x(k+1)$ denotes the successor state.

The state and input are subject to the constraints
\begin{eqnarray}
x(k) \in \mathbb{X} \subseteq \mathbb{R}^n,~~ u(k) \in  \mathbb{U} \subseteq \mathbb{R}^m.
\end{eqnarray}

Additionally, the disturbance satisfies the condition of 
\begin{eqnarray} \label{distB}
w(k) \in  \mathbb{W} \subseteq \mathbb{R}^q.
\end{eqnarray}

The objective of this paper is to design a control input $u(k)$ for the disturbed nonlinear system (1) to drive the state $x(k)$ to the neighbourhood of the origin with the proposed MPC.
 
\emph{Assumption 1.} The constraint sets $\mathbb{X}$, $\mathbb{U}$, and $\mathbb{W}$ are PC-sets.

\emph{Assumption 2.} The function $f(x(k),u(k),w(k))$ is continuous for all $x(k) \in \mathbb{X}$, $u(k) \in \mathbb{U}$, $w(k) \in \mathbb{W}$, and satisfies $\left\|  f(x(k),u(k),w(k))-f(x(k),u(k),0)\right\| \le \delta_1 (\left\|w(k)\right\|)$ for a $\mathcal{K}_{\infty}$-function $\delta_1(\cdot)$.

\emph{Assumption 3.}
1) The full state $x(k)$ at current time $k$ is available and known exactly. \\
2) An $N$-step prediction of future disturbance at current time $k$ is available
\begin{eqnarray}
\bold w (k) \buildrel \Delta \over = \left\{w(k|k),w(k+1|k),\cdots,w(k+N-1|k) \right\}.
\end{eqnarray}

\emph{Remark 1.}  
The prediction of the future disturbance or preview may be available by sensor measurements in some engineering systems, e.g.  wave energy converter (WEC) system \cite{18}, autonomous vehicle system \cite{19}, and so on. Furthermore, the disturbance preview information also can be obtained by soft sensors, such as disturbance observer techniques \cite{14}.

\subsection{Definitions and Lemmas}

\emph{Definition 1. (Robust Positively Invariant Set \cite{21})} A set $\Omega \subseteq \mathbb{R}^n $ is called a robust positively invariant (RPI) set for nonlinear system (1), if for all initial state $x_0 \in \Omega$, the state at time $k$ satisfies $x(k,x_0,w) \in \Omega $ for all $w \in \mathbb{W}$.

\emph{Definition 2. (ISS-Lyapunov function \cite{22})} A continuous function $V(\cdot):\mathbb{R}^n \to \mathbb{R}_{\ge 0}$ is called an ISS-Lyapunov function for nonlinear system (1) if the following holds:

1) There exist $\mathcal{K}_{\infty}$-functions $\alpha_1(\cdot)$ and $\alpha_2(\cdot)$ such that
\begin{eqnarray}
\alpha_1(\left\|x \right\|) \le V(x) \le \alpha_2(\left\|x \right\|), \forall x \in \mathbb{R}^n.
\end{eqnarray}

2) There exist a $\mathcal{K}_{\infty}$-function $\alpha_3(\cdot)$ and a $\mathcal{K}$-function $\rho(\cdot)$ such that
\begin{eqnarray}
V(f(x,w)) - V(x)    \le - \alpha_3(\left\|x \right\|) + \rho (\left\|w \right\|),
\end{eqnarray}
for all $x \in \mathbb{R}^n $ and $w \in \mathbb{R}^q $.

\emph{Lemma 1\cite{22}.} If the nonlinear system (1) admits a continuous ISS Lyapunov function $V(\cdot)$, then the system is ISS.

\section{MPC Framework With Preview Information of Disturbance}
To achieve the control objective, an MPC framework with preview information of disturbance is proposed in this section.

\subsection{Design of MPC with disturbance preview information}
Consider the $N$-step prediction of future disturbance at time instant $k$: $\bold w (k) = \left\{w(k|k),\cdots,w(k+N-1|k) \right\}$. The dynamics of the disturbance preview information can be described by
\begin{eqnarray}
\begin{array}{l}
\bold w(k + 1) = {A_w} \bold w(k) + {B_w}{w_0}(k),
\end{array}
\end{eqnarray}
where $A_w=\left[ {\begin{array}{*{20}{c}}
0&1&0& \cdots &0\\
 \vdots & \vdots & \vdots & \ddots & \vdots \\
0&0&0& \cdots &1\\
0&0&0& \cdots &0
\end{array}} \right]$, $B_w=\left[ {\begin{array}{*{20}{c}}
0\\
 \vdots \\
0\\
{1}
\end{array}} \right]$.
$\bold w(k + 1) $ is the $N$-step prediction of future disturbance at time instant $k+1$. $\omega_0(k)$ is the first disturbance outside the horizon so it is unknown at time $k$ but satisfies (\ref{distB}), i.e. $\omega_0(k) \in \mathbb{W}$.   

Considering the original nonlinear system (1) and the disturbance preview dynamics (7), the augmented system is given by
\begin{eqnarray}
\begin{array}{l}
{z(k + 1)} = F(z(k),u(k)) + Gw_0(k),
\end{array}
\end{eqnarray}
where $z(k) =\left[ {\begin{array}{*{20}{c}}
{x^T(k)}&{\bold w^T(k)}
\end{array}} \right]^T$, $F(z(k),u(k))=\left[ {\begin{array}{*{20}{c}}
f(x(k),u(k),\bold w(k))\\
{A_w} \bold w(k) 
\end{array}} \right] $, and $G=\left[ {\begin{array}{*{20}{c}}
0\\
{{B_w}}
\end{array}} \right]$.

Then, the MPC problem for the augmented system (8) at time $k$ with prediction horizon $N$ is formulated by
\begin{eqnarray}
\begin{array}{l}
V^0_N(z(k)) = \mathop {\min }\limits_{\bold u(k) \in \mathcal{U}_N}  V_N(x(k),\bold{u}(k),\bold{w}(k)) ,
\end{array}
\end{eqnarray}
subject to
\begin{small}
\begin{eqnarray}
\left\{ \begin{array}{l}
x(k|k)=x(k), \\
x(k+i+1|k)=  f(x(k+i|k),u(k+i|k),w(k+i|k)), \\
x(k+i|k) \in \mathbb{X},~u(k+i|k) \in \mathbb{U},~x(k+N|k) \in \mathbb{X}_f,
\end{array} \right.
\end{eqnarray}
\end{small}
with $i=0,1,\cdots,N-1$. $\mathbb{X}_f$ is the terminal constraint set.

The cost function in this MPC problem considering the preview information of disturbance is given by
\begin{eqnarray}
\begin{array}{l}
V_N(z(k))  \buildrel \Delta \over = \sum\limits_{i=0}^{N-1} {l \left( {x(k+i|k),u(k+i|k),w(k+i|k)} \right) } \\
~~~~~~~~~~~~~~~+ V_f \left(x(k+N|k)\right),
\end{array}
\end{eqnarray}
where $l \left( {\cdot,\cdot,\cdot} \right): \mathbb{R}^n \times  \mathbb{R}^m \times \mathbb{R}^q \to \mathbb{R}_{\ge 0} $ denotes the stage cost function to evaluate the control performance, energy loss, as well as the penalty of disturbance for the control process. $V_f \left( \cdot \right):\mathbb{R}^n  \to \mathbb{R}_{\ge 0} $ denotes the terminal cost to penalize the terminal state. To simplify the discussion, the stage and terminal cost functions are considered in the quadratic form: $l \left( {x(k+i|k),u(k+i|k),w(k+i|k)} \right) =\left\| x(k+i|k) \right\|^2_Q+ \left\| u(k+i|k) \right\|^2_R + \left\| w(k+i|k) \right\|^2_S $, and $V_f \left( {x(k+N|k)} \right) = \left\| x(k+N|k) \right\|^2_P  $ with symmetric weighting matrices $Q \ge 0$, $R>0$, $S>0$, and $P \ge 0$. It is possible to extend the main results in this paper to the stage cost in a non-quadratic form as in the economic MPC or other applications with an appropriate modifications and technical development. 

\emph{Remark 2.} The prediction model in optimisation problem (10) considers the preview information of disturbance, which makes it possible to take full advantage of the disturbance to improve control performance. The proposed stage cost  depends on not only the system state and input, but also the disturbance, which is more general than other MPC algorithms with preview, i.e. \cite{20}. This feature is vital in establishing the stability and recursive feasibility of our proposed algorithms.  

\subsection{Design of terminal conditions}
In this section, the terminal ingredients $V_f(\cdot)$ and $\mathbb{X}_f$ of the proposed MPC algorithm will be developed to guarantee the recursive feasibility and stability of the control problem.

To proceed, we first make the following assumption for the nonlinear system and optimisation problem.

\emph{Assumption 4.} For all $x \in \mathbb{X}_f$, there exists a $u_f \in \mathbb{U}$ and a $\mathcal{K}_{\infty}$-function $ \sigma_1(\left\|x \right\|) $ such that
\begin{eqnarray}
\begin{array}{l}
f(x,u_f,w) \in \mathbb{X}_f, ~~\forall w \in \mathbb{W},
\end{array}
\end{eqnarray}
and
\begin{eqnarray}
\begin{array}{l}
V_f(f(x,u_f,0)) \le V_f(x)  - l (x,u_f,0) - \sigma_1(\left\|x \right\|).
\end{array}
\end{eqnarray}

Considering the continuity of the terminal cost $V_f(\cdot)$ and Assumption 2, we have
\begin{eqnarray}
\begin{array}{l}
~~~\left\|V_f(f(x,u_f,w)) - V_f (f(x,u_f,0)) \right\| \\
\le  \delta_2 (\left\|f(x,u_f,w)-f(x,u_f,0) \right\|) \\
\le  \delta_2 \circ \delta_1 (\left\|w \right\|) \\
\le  \delta_3(\left\|w \right\|),
\end{array}
\end{eqnarray}
with $\mathcal{K}_{\infty}$-functions $\delta_1(\cdot)$, $\delta_2(\cdot)$ and $\delta_3(\cdot)$.

Furthermore, it follows from (13) and (14) that
\begin{eqnarray}
\begin{array}{l}
V_f(f(x,u_f,w))  \le V_f(f(x,u_f,0))  + \delta_3(\left\| w \right\|) \\
~~~~~~~~~~~\le  V_f(x)  - l (x,u_f,0) -\sigma_1(\left\|x \right\|) + \delta_3(\left\|w \right\|) \\
\end{array}
\end{eqnarray}
for all $w \in \mathbb{W}$.

\emph{Lemma 2.} The closed-loop terminal system $x^+=f(x,u_f,w)$ is ISS.

\emph{Proof:} Since the terminal cost is defined by $V_f (x) = \left\| x \right\|^2_P  $, we have $\lambda_{\mathrm{min}}(P)  \left\| x \right\|^2 \le V_f (x)   \le \lambda_{\mathrm{max}}(P)  \left\| x \right\|^2$. In addition, according to (15), we have $V_f(f(x,u_f,w))  -  V_f(x)  \le - l (x,u_f,0) -\sigma_1(\left\|x \right\|) + \delta_3(\left\|w \right\|)$.  Since the stage cost is defined by $l(x,u_f,w) = \left\| x \right\|^2_Q + \left\| u_f \right\|^2_R + \left\| w \right\|^2_S  $, we have $V_f(f(x,u_f,w))  -  V_f(x)  \le  -\sigma(\left\|x \right\|) + \delta_3(\left\|w\right\|)$ for the $\mathcal{K}_{\infty}$-function $ \sigma(\left\|x\right\|) = \left\| x \right\|^2_Q +\sigma_1(\left\|x\right\|) $. It follows from Definition 2 that the terminal cost $V_f (x)$ is an ISS-Lyapunov function for the closed-loop terminal system $x^+=f(x,u_f,w)$. Therefore, according to Lemma 1, the closed-loop terminal system $x^+=f(x,u_f,w)$ is ISS.

\emph{Remark 3.} It follows from Definition 2 and Lemma 2 that the $\mathcal{K}_{\infty}$-functions $\alpha_1(\left\| x \right\|) =\lambda_{\mathrm{min}}(P)  \left\| x \right\|^2 $, $\alpha_2(\left\|x \right\|) =\lambda_{\mathrm{max}}(P)  \left\| x \right\|^2 $, $\alpha_3(\left\|x \right\|) =\left\| x \right\|^2_Q +\sigma_1(\left\|x\right\|)  $, and $\rho(\left\|w \right\|) =\delta_3(\left\|w \right\|)$. According to \cite{22}, the states of system $x^+=f(x,u_f,w)$  will converge to an invariant set $\Omega(\beta)=\{x|V_f(x)\le \beta \}$ for $\beta=\alpha_2 \circ \alpha_3^{-1}  \circ \rho(\left\|w \right\|)$.

\subsection{Theoretical analysis}
In this subsection, theoretical analysis for the recursive feasibility and stability of the proposed MPC with preview information of disturbance is given.

\emph{Theorem 1.}  Suppose that Assumptions 1-4 are satisfied, and the optimisation problem (9)-(10) is feasible at time $k=0$. Then the proposed MPC algorithm is recursively feasible under the prescribed disturbance.

\emph{Proof:} First of all, assume that the proposed MPC algorithm is feasible at time instant $k \in \mathbb{I}_{\ge 0}$. The optimal control sequence at time instant $k$ is denoted by
\begin{eqnarray}
\begin{array}{l}
\bold u^* (k)  = \left\{u^*(k|k),  \cdots, u^*(k+N-1|k) \right\},
\end{array}
\end{eqnarray}
and the corresponding optimal state sequence is given by
\begin{eqnarray}
\begin{array}{l}
\bold x^* (k)  = \left\{x^*(k|k),\cdots,x^*(k+N|k) \right\},
\end{array}
\end{eqnarray}
where the terminal state satisfies $x^*(k+N|k) \in \mathbb{X}_f$.

In what follows, a control sequence at time instant $k+1$ will be constructed based on the current optimal sequence, which is denoted by
\begin{eqnarray}
\tilde {\bold {u}} (k+1) = \left\{ \begin{array}{l}
u^*(k+i|k),  ~~~~~~~~~~~~~~i \in \mathbb{I}_{[1,N-1]}, \\
u_f(x^*(k+N|k)), ~~~~~~~i =N.
\end{array} \right.
\end{eqnarray}

Under the constructed control sequence (18) at time instant $k+1$, the corresponding state trajectory can be obtained by $x(k+i|k+1) = x^*(k+i|k) \in \mathbb{X}$ for $i \in \mathbb{I}_{[1,N-1]}$, and $x(k+N|k+1)=x^*(k+N|k)  \in \mathbb{X}_f$. According to Assumption 4, there exists a control 
\begin{equation}
u(k+N|k+1)=u_f(x^*(k+N|k)) \in \mathbb{U} ,  \nonumber
\end{equation}
such that the resulting state
\begin{equation}
x(k+N+1|k+1) \in \mathbb{X}_f.  \nonumber
\end{equation}

Therefore, according to the above analysis, the constructed control sequence (18) is a feasible solution of the proposed MPC algorithm (9)-(10) at time instant $k+1$, which means that the proposed MPC algorithm is recursively feasible. 

In what follows, we will give the stability analysis of the proposed MPC algorithm.

\emph{Theorem 2.}  Suppose that Assumptions 1-4 are satisfied. Then the closed-loop system under the proposed MPC algorithm (9)-(10) is input-to-state stable (ISS) under the described  disturbance.

\emph{Proof:} To show the ISS of the closed-loop system, we define a Lyapunov function candidate as $V^0_N(z(k))=  \sum\limits_{i=0}^{N-1} {l \left( {x^*(k+i|k),u^*(k+i|k),w(k+i|k)} \right) } + V_f \left(x^*(k+N|k)\right)$, at time instant $k$.

In the first step, we will show that there exists a lower bound for the Lyapunov function $V^0_N(z(k))$. It is obviously that $V^0_N(z(k)) \ge l \left( {x^*(k),u^*(k),w(k|k)} \right) +\sum\nolimits_{i=1}^{N-1} {l(0,0,w(k+i|k))} $. Since the stage cost is quadratic, there exists a $\mathcal{K}_{\infty}$-function $c_1(\cdot)$ such that $l \left( {x^*(k),u^*(k),w(k|k)} \right) +\sum\nolimits_{i=1}^{N-1} {l(0,0,w(k+i|k))}  \ge c_1(\left\|z(k) \right\|)$. Therefore, the following condition holds
\begin{eqnarray}
\begin{array}{l}
V^0_N(z(k)) \ge c_1(\left\|z(k) \right\|),
\end{array}
\end{eqnarray}
for all $x \in \mathcal{X}_N$.

In the second step, we will also show that there exists an upper bound for the Lyapunov function $V^0_N(z(k))$. Considering the continuity of the cost function $V^0_N(z(k)) = V^0_N(x(k), \bold w(k))$, we have $\left\|V^0_N(x(k), \bold w(k)) - V^0_N(x(k), 0) \right\| \\ \le \delta_5 (\left\|\bold w(k) \right\|)$ with a $\mathcal{K}_{\infty}$-function $\delta_5(\cdot)$. Thus, the condition $V^0_N (x(k), \bold w(k))  \le  V^0_N(x(k), 0)  + \delta_5 (\left\|\bold w(k) \right\|) $ holds. According to \cite{1}, we have $V^0_N(x(k), 0) \le  c_2(\left\|x(k) \right\|)$ with a $\mathcal{K}_{\infty}$-function $c_2(\cdot)$. Thus, there exists a $\mathcal{K}_{\infty}$-function $c_3(\cdot)$ such that $V^0_N(z(k)) \le c_2(\left\|x(k) \right\|) + \delta_5(\left\|\bold w(k) \right\|)  \le c_2(\left\|z(k) \right\|) + \delta_5(\left\|z(k) \right\|) \le c_3(\left\|z(k)\right\|) $ holds. Therefore, the following condition holds
\begin{eqnarray}
\begin{array}{l}
V^0_N(z(k)) \le c_3(\left\|z(k) \right\|),
\end{array}
\end{eqnarray}
for all $x \in \mathcal{X}_N$.

In the third step,  we will show the descent property of Lyapunov function $V^0_N(z(k))$.

Considering the condition (15) and the constructed feasible control sequence (18), we have
\begin{eqnarray}
\begin{array}{l}
~~~V^0_N(z(k+1))  - V^0_N(z(k)) \\
\le V_N(x(k+1),\tilde {\bold u}(k+1), \bold w(k+1))  - V^0_N(z(k)) \\
=   l (x^*(k+N|k),u(k+N|k+1),w(k+N|k+1)) \\
~~  + V_f(x(k+N+1|k+1)) -V_f(x^*(k+N|k))   \\
~~   -l(x^*(k),u^*(k),w(k|k)) \\
\le  - \sigma_1(\left\|x^*(k+N|k) \right\|) +\delta(\left\|w(k+N|k+1) \right\|)  \\
~~~ -l(x^*(k),u^*(k),w(k|k)),
\end{array}
\end{eqnarray}
for a $\mathcal{K}_{\infty}$-function $\delta(\left\|w(k+N|k+1) \right\|) = l(0,0,w(k+N|k+1)) + \delta_3(\left\|w(k+N|k+1) \right\|) $.

It follows from the condition (19) that 
\begin{eqnarray}
\begin{array}{l}
~~~V^0_N(z(k+1))  - V^0_N(z(k)) \\
\le - \sigma_1(\left\|x^*(k+N|k) \right\|) +\delta(\left\|w(k+N|k+1)\right\|)  \\
~~~ - c_1(\left\|z(k)\right\|) + \sum\nolimits_{i=1}^{N-1} {l(0,0,w(k+i|k))} \\
\le - \sigma_1(\left\|x^*(k+N|k) \right\|) + \bar \rho(\left\|\bold w(k+1)\right\|) \\
~~~ - c_1(\left\|z(k)\right\|). 
\end{array}
\end{eqnarray}
for a $\mathcal{K}_{\infty}$-function $\bar \rho(\left\|\bold w(k+1)\right\|) =  \delta(\left\|w(k+N|k+1)\right\|) +\sum\nolimits_{i=1}^{N-1} {l(0,0,w(k+i|k))}$. According to the relationship (22), if $x^*(k+N|k)$ is outside the region $\Omega=\{x(k+N|k)| \sigma_1(\left\|x(k+N|k) \right\|) \le \bar \rho(\left\|\bold w(k+1)\right\|) \}$, then $V^0_N(z(k+1))  < V^0_N(z(k))$ holds, which means the state $z(k)$ is robustly asymptotically stable until the terminal state $x^*(k+N|k)$ converges to the stability domain $\Omega$.  Therefore, the closed-loop system under the proposed MPC is input-to-state stable (ISS) with respect to the disturbance $w \in \mathbb{W}$.

\section{The implementation issues}
In this section, implementation issues about the proposed MPC algorithm will be discussed to facilitate its practical applications.

\subsection{The implementation of the MPC algorithm}
First of all, the  implementation of the proposed MPC algorithm is given as follows.

\textbf{Offline:} \\
\noindent \textbf{1.}  Select the weight matrices $Q$, $R$ and $S$ for the stage cost $l(x,u,w)$. \\
\noindent \textbf{2.}  Check the satisfaction of Assumptions 1-3.\\
\noindent \textbf{3.}  Select the weight matrix $P$ of the terminal cost $V_f(x)$ and the $\mathcal{K}_{\infty}$-function $ \sigma_1(\left\|x \right\|) $ to check the satisfaction of Assumption 4.\\
\noindent \textbf{4.}  Compute the terminal constraint set $\mathbb{X}_f$.

\textbf{Online:} \\
\noindent \textbf{Step1.}  At time instant $k$, measure the current state $x(k)$, and update the preview information of disturbance
\begin{equation}
\bold w (k) = \left\{w(k|k),w(k+1|k),\cdots,w(k+N-1|k) \right\} . \nonumber
\end{equation}
\noindent \textbf{Step2.}  Solve the optimisation problem (9)-(10), and obtain the optimal solution $\bold u^*(k)$ and $\bold x^*(k)$.\\
\noindent \textbf{Step3.}  Apply the optimal control $u(k)=u^*(k|k)$ to the actual nonlinear system (1).\\
\noindent \textbf{Step4.}  Set $k=k+1$, and go back to Step 1.

\subsection{The calculation of the terminal conditions}
This subsection is devoted to calculating the terminal constraint set $\mathbb{X}_f$ and terminal cost $V_f(x)$ of the proposed MPC algorithm.

It should be noted that Assumptions 1-3 in the proposed MPC algorithm are almost the same with that of the existing MPC framework with stability guarantee. However, there exists significant difference about the terminal cost in Assumption 4. The important stability-guarantee condition (13) contains an extra $\mathcal{K}_{\infty}$-function $ \sigma_1(\left\|x\right\|) $, which provides an extra freedom for the control system to obtain a larger stability margin to accommodate the influence of the disturbance and strengthen its disturbance-rejection capability.

We consider the Jacobian linearisation of the nonlinear system (1) at the origin
\begin{eqnarray}
x(k+1) =  Ax(k) + Bu(k) + B_w w(k),
\end{eqnarray}
where $A = {\left. {\frac{{\partial f(x,u,w)}}{{\partial x}}} \right|_{(0,0,0)}}$, $B = {\left. {\frac{{\partial f(x,u,w)}}{{\partial u}}} \right|_{(0,0,0)}}$, $B_w = {\left. {\frac{{\partial f(x,u,w)}}{{\partial w}}} \right|_{(0,0,0)}}$. Select the stage cost and terminal cost as $l \left( {x,u,w} \right) =\left\| x \right\|^2_Q+ \left\| u \right\|^2_R + \left\| w \right\|^2_S $ and $V_f \left( {x} \right) = \left\| x \right\|^2_P  $, respectively. The $\mathcal{K}_{\infty}$-function $ \sigma_1(\left\|x \right\|) $ in Assumption 4 is also defined in the quadratic form $ \sigma_1(\left\|x \right\|) = \left\| x \right\|^2_\Delta$ with the symmetric weighting matrix $\Delta \ge 0$.

We assume that the linearised system (23) is stabilisable. Therefore, there exists a linear state feedback control $u_f =Kx$ such that $A_K=A+BK$ is strictly stable.

First of all, we will propose a procedure to find a terminal constraint set $\mathbb{X}_f$ to satisfy the condition (12) in Assumption 4 based on the linearised system (23).

To achieve more accurate approximation of the RPI set, the following polytopic set $\mathcal{P}$ is selected to approximate the terminal constraint set $\mathbb{X}_f$, rather than ellipsoidal set,
\begin{eqnarray}
\begin{array}{l}
\mathcal{P}  \buildrel \Delta \over =  \{x| -\varepsilon \le H x \le \varepsilon  \},
\end{array}
\end{eqnarray}
where $0<\varepsilon \in \mathbb{R}^p$, $H \in \mathbb{R}^{p \times n}$ and $p \ge n$. $p$ can be selected based on the required accuracy. If $p > n$, the polytope is a full-complexity polytope, which is used in this paper. By using the designed algorithm in \cite{23}, both the terminal set $\mathbb{X}_f$ and control gain $K$ can be calculated for the proposed MPC.

In what follows, we will select the matrices $\Delta$ and $P$ to satisfy the condition (13). 

The condition (13) can be rewritten by
\begin{eqnarray} \nonumber
\begin{array}{l}
x^T(A+BK)^T P (A+BK) x \le x^T P x \\
~~~~~~~~~~~~~~~~~~~~~~~~~~~~~ -  x^T (Q+K^TRK) x -  x^T \Delta x ,
\end{array}
\end{eqnarray}
$\Leftrightarrow$
\begin{eqnarray}
\begin{array}{l}
P \ge (A+BK)^T P (A+BK)  +  (Q+K^TRK)  +  \Delta.
\end{array}
\end{eqnarray}

In order to meet the condition (25), we select a positive definite matrix $\Delta$. Then, let the matrix $P$ be defined by the following Lyapunov equation
\begin{eqnarray}
\begin{array}{l}
P = (A+BK)^T P (A+BK)  +  (Q+K^TRK + \lambda \Delta),
\end{array}
\end{eqnarray}
for some $\lambda \ge1$. Since $A+BK$ is stable and $Q+K^TRK + \lambda \Delta$ is positive definite, $P$ is positive definite. Therefore, the condition (25), and hence (13), is satisfied.

\section{Illustrative example}
Consider a linear, discrete-time, constrained system subject to additive bounded disturbance
\begin{eqnarray}
\begin{array}{l}
x^+ =\left[ {\begin{array}{*{20}{c}}
1&1\\
0&1
\end{array}} \right] x
+\left[ {\begin{array}{*{20}{c}}
0.5\\
1
\end{array}} \right] u
+\left[ {\begin{array}{*{20}{c}}
1&0\\
0&1
\end{array}} \right] w.
\end{array}
\end{eqnarray}

The state, input and disturbance constraints are given by $\mathbb{X}=\{x \in \mathbb{R}^2: ||x||_{\infty} \le 1 \}$, $\mathbb{U}=\{u \in \mathbb{R}: ||u||_{\infty} \le 1 \}$, and $\mathbb{W}=\{w\in \mathbb{R}^2: ||w||_{\infty} \le 0.1\}$.

The weighting matrices of the stage cost are selected by $Q = \rm{diag} \{10, 1\}$, $R=1$ and $S=I$. The terminal set $\mathbb{X}_f$ and control gain $K$ are obtained with the algorithm in \cite{23}. The weighting matrix $P$ is calculated by (26) with $\Delta=I$ and $\lambda=2$.

Fig. 1 depicts the terminal set $\mathbb{X}_f$ of the proposed MPC algorithm in the phase plane. Fig. 2 gives the trajectories under three MPC algorithms: proposed MPC, nominal MPC, and discrete-time form of DRMPC \cite{17} that compensates the disturbance by feedforward technique and then attenuates the residual disturbance by MPC optimisation problem. It shows that the trajectory under the proposed MPC algorithm can converge to a smallest region of the origin among these three algorithms. Therefore, we have that the proposed MPC algorithm possesses the strongest disturbance-rejection capability. Furthermore, the actual running costs, calculated by summing all the stage costs without disturbance over time, under the three MPC algorithms are given in Table 1, which shows that the proposed MPC also achieves a smallest cost value. In summary, the consideration of the disturbance preview information in MPC optimisation problem makes it possible to make full use of the disturbance.
\begin{figure}
\centering{\includegraphics[scale=0.42]{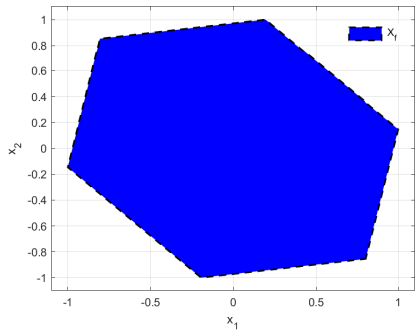}}
\caption{The terminal region $\mathbb{X}_f$}
\end{figure}
\begin{figure}
\centering{\includegraphics[scale=0.42]{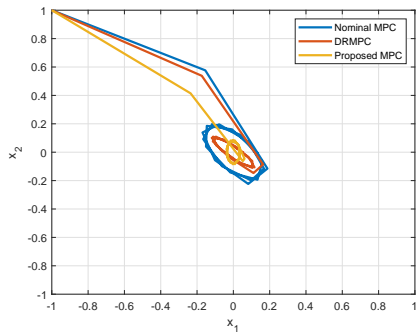}}
\caption{The trajectories under three MPC algorithms}
\end{figure}
\begin{table}
\caption{The Actual Running Costs Under Three MPC Algorithms}
\label{table}
\setlength{\tabcolsep}{3pt}
\begin{tabular}{|p{68pt}|p{52pt}|p{52pt}|p{52pt}|}
\hline
Algorithms&Nominal MPC&DRMPC &Proposed MPC\\
\hline
Actual running costs & 15.900& 13.883&12.874 \\
\hline
\end{tabular}
\label{tab1}
\end{table}

\section{Conclusions}
A novel model predictive control algorithm with preview information of disturbance for nonlinear systems is proposed in this paper. It is assumed that the disturbance is known within the prediction horizon and unknown out the horizon. By considering reference (tasks) and the predicted disturbance within the horizon in a single optimisation framework, it can make full use of disturbance information to improve the control performance. Recursive feasibility and input-to-state stability of the proposed MPC algorithm under the described disturbance are established by carefully constructing a terminal condition and a terminal cost. In this study,  it is assumed that the disturbance is exactly known. But in practice, the disturbance preview may be corrupted by sensor noise if it is measured, or have estimation error if it is estimated by a disturbance observer technique. We will conduct our future research in this direction, particularly incorporating the estimation dynamics of a disturbance observer.



\begin{thebibliography}{00}
\bibitem{1} J. B. Rawlings and D. Q. Mayne, \emph{Model Predictive Control: Theory and Design.} Madison, WI, USA: Nob Hill Publishing, 2009.

\bibitem{2} D. Q. Mayne, J. B. Rawlings, C. V. Rao, and P. O. M. Scokaert, ``Constrained model predictive control: stability and optimality,'' \emph{Automatica}, vol. 36, pp. 789--814, Jun. 2000.

\bibitem{3} W.-H. Chen, D. J. Ballance, and P. J. Gawthrop, ``Optimal control of nonlinear systems: a predictive control approach,'' \emph{Automatica}, vol. 39, pp. 633--641, Apr. 2003.

\bibitem{4} B. Kouvaritakis and M. Cannon, \emph{Model Predictive Control: Classical, Robust and Stochastic.} New York, NY, USA: Springer, 2016.

\bibitem{5} G. Grimm, M. J. Messina, S. E. Tuna, and A. R. Teel, ``Nominally robust model predictive control with state constraints,'' \emph{IEEE Trans. Autom. Control}, vol. 52, pp. 1856--1870, Oct. 2007.

\bibitem{6} D. Q. Maynea, M. M. Seronb, and S.V. Rakovic, ``Robust model predictive control of constrained linear systems with bounded disturbances,'' \emph{Automatica}, vol. 41, pp. 219--224, Feb. 2005.

\bibitem{7} W. Langson, L. Chryssochoos, S.V. Rakovic, and D.Q. Mayne, ``Robust model predictive control using tubes,'' \emph{Automatica}, vol. 40, pp. 125--133, Jan. 2004.

\bibitem{8} J. Fleming, B. Kouvaritakis, and M. Cannon, ``Robust model predictive control of constrained linear systems with bounded disturbances,'' \emph{IEEE Trans. Autom. Control}, vol. 60, pp. 1087--1092, Feb. 2015.

\bibitem{9} P. O. M. Scokaert and D. Q. Mayne, ``Min-max feedback model predictive control for constrained linear systems,'' \emph{IEEE Trans. Autom. Control}, vol. 43, pp. 1136--1142, Aug. 1998.

\bibitem{10} D. Limon, T. Alamo, F. Salas, and E. F. Camacho, ``Input to state stability of min–max MPC controllers for nonlinear systems with bounded uncertainties,'' \emph{Automatica}, vol. 42, pp. 797--803, May. 2006.

\bibitem{11} A. Mesbah, ``Stochastic model predictive control: an overview and perspectives for future research,'' \emph{IEEE Control Syst. Mag.}, vol. 36, pp. 30--44, Dec. 2016.

\bibitem{12} Z.-Q. Sun, V. Rostampour, and M. Cao, ``Self-triggered stochastic MPC for linear systems with disturbances,'' \emph{IEEE Control Syst. Lett.}, vol. 3, pp. 787--792, Oct. 2019.

\bibitem{13} Z.-G. Wang, J. Zhan, C.-G. Duan, X. Guan, P.-P. Lu, and K. Yang, ``A review of vehicle detection techniques for intelligent vehicles,'' \emph{IEEE Trans. Neural Netw. Learn. Syst.}, Early Access, DOI: 10.1109/TNNLS.2021.3128968.

\bibitem{14} W.-H. Chen, J. Yang, L. Guo, and S.-H. Li, ``Disturbance-observer-based control and related methods--An overview,'' \emph{IEEE Trans. Ind. Electron.}, vol. 63, pp. 1083--1095, Feb. 2016.

\bibitem{15} C.-J. Liu, W.-H. Chen, and J. Andrews, ``Tracking control of small-scale helicopters using explicit nonlinear MPC augmented with disturbance observers,'' \emph{Control Eng. Pract.}, vol. 20, pp. 258--268, Mar. 2012.

\bibitem{16} J. Yang, W.-X. Zheng, S.-H. Li, B. Wu, and M. Cheng, ``Design of a prediction accuracy enhanced continuous-time MPC for disturbed systems via a disturbance observer,'' \emph{IEEE Trans. Ind. Electron.}, vol. 62, pp. 5807--5816, Sep. 2015.

\bibitem{17} H.-H. Xie, L. Dai, Y.-C Lu, and Y.-Q Xia, ``Disturbance rejection MPC framework for input-affine nonlinear systems,'' \emph{IEEE Trans. Autom. Control}, to be published.

\bibitem{18} S.-Y. Zhan and G. Li, ``Linear optimal noncausal control of wave energy converters,'' \emph{IEEE Trans. Control Syst. Technol.}, vol. 27, pp. 1526--1536, Jul. 2019.

\bibitem{19} S.-B. Xu and H. Peng, ``Design, analysis, and experiments of preview path tracking control for autonomous vehicles,'' \emph{IEEE Trans. Intell. Transp. Syst.}, vol. 21, pp. 48--58, Jan. 2020.

\bibitem{20} P. R. B. Monasterios and P. A. Trodden, ``Model predictive control of linear systems with preview information: feasibility, stability, and inherent robustness,'' \emph{IEEE Trans. Autom. Control}, vol. 64, pp. 3831--3838, Sep. 2019.

\bibitem{21} F. Blanchini, ``Set invariance in control,'' \emph{Automatica}, vol. 35, pp. 1747--1767, Nov. 1999.

\bibitem{22} Z.-P. Jiang and Y. Wang, ``Input-to-state stability for discrete-time nonlinear systems,'' \emph{Automatica}, vol. 37, pp. 857--869, Jun. 2001.

\bibitem{23} C.-Y. Liu, F. Tahir, and I. M. Jaimoukha, ``Full-complexity polytopic robust control invariant sets for uncertain linear discrete time systems,'' \emph{Int J Robust Nonlinear Control}, vol. 29, pp. 3587--3605, Jul. 2019.

\end{thebibliography}
\end{document}